\documentclass[aps,prb,reprint,superscriptaddress]{revtex4-1}
\usepackage{graphicx,amssymb,amsmath,amsfonts,verbatim,color,xr}

\bibpunct{[}{]}{,}{n}{}{}

\begin{document}
\bibliographystyle{apsrev}

\title{Vortex crossing and trapping in doubly connected mesoscopic loops of a single-crystal type II superconductor}

\author{Shaun A. Mills}
\affiliation{Department of Physics and Materials Research Institute, The Pennsylvania State University, University Park, PA 16802, USA}

\author{Chenyi Shen}
\author{Zhuan Xu}

\affiliation{Department of Physics, Zhejiang University, Hangzhou 310027, China}
\affiliation{Collaborative Innovation Center of Advanced Microstructures, Nanjing 210093, China}

\author{Ying Liu}
\email{yxl15@psu.edu}

\affiliation{Department of Physics and Materials Research Institute, The Pennsylvania State University, University Park, PA 16802, USA}

\affiliation{Collaborative Innovation Center of Advanced Microstructures, Nanjing 210093, China}

\affiliation{Department of Physics and Astronomy and Key Laboratory of Artificial Structures and Quantum Control 
(Ministry of Education), Shanghai Jiao Tong University, Shanghai 200240, China}

\begin{abstract}
Numerical calculations on a mesoscopic ring of a type II superconductor in the London limit suggest that an Abrikosov vortex can be trapped in such a structure above a critical magnetic field and generate a phase shift in the magnetoresistance oscillations. We prepared submicron-sized superconducting loops of single-crystal, type II superconductor NbSe$_2$ and measured magnetoresistance oscillations resulting from vortices crossing the loops. The free energy barrier for vortex crossing determines the crossing rate and is periodically modulated by the external magnetic flux threading the loop. We demonstrated experimentally that the crossing of vortices can be directed at a pair of constrictions in the loop, leading to more pronounced magnetoresistance oscillations than those in a uniform ring. The vortex trapping in both a simple ring and a ring featuring two constrictions was found to result in a phase shift in the magnetoresistance oscillations as predicted in the numerical calculations. The controlled crossing and trapping of vortices demonstrated in our NbSe$_2$ devices provide a starting point for the manipulation of individual Abrikosov vortices, which is useful for future technologies. 
\end{abstract}

\maketitle

Vortex motion in a type II superconductor is an important problem to consider for increasing the critical current density and upper critical field of superconducting materials, both of which are crucial for applications of superconductors such as electromagnets and magnetic levitation. Manipulating the motion of Abrikosov vortices is also of fundamental interest. It is known that a moving Abrikosov vortex is subjected to various fundamental forces including damping, pinning, boundary image, and transverse (Lorentz and Magnus) forces\cite{HuebenerBook}. None of these forces influencing vortex motion are fully understood. Indeed, even the effective vortex mass upon which the forces act remains a subject of controversy\cite{SuhlVortexMass,ChudnovskyVortexMass,FilVortexMass,GolubchikVortexMass}. Additionally, Abrikosov vortices in conventional type II superconductors have long been a model system for motion in soft matter\cite{BlatterRMPVortices}, and controlled vortex manipulation in superconducting devices has potential application in rectifiers\cite{VillegasRectifier}, superconducting logic circuits\cite{HastingsVortexRatchet}, and hybrid superconductor/dilute magnetic semiconductor spintronic systems\cite{BerciuVortexSpintronics}.

In order to manipulate and detect the motion of individual Abrikosov vortices, a scanning superconducting quantum interference device (SQUID) and magnetic force microscope were employed on planar films of YBa$_3$Cu$_2$O$_{6.354}$\cite{GardnerYBCOVorManip} and Nb\cite{StraverNbVorManip}, respectively. On the other hand, little work has been done on the manipulation and detection of vortex motion in mesoscopic superconductors, which are more relevant for technological applications than planar films. Static few vortex states in mesoscopic superconductors have been studied previously both theoretically\cite{DeoMesoscopicDisk,BruyndoncxCrossover,ChibotaruAlTriangle,KoganLondon,PekkerBlockade} and experimentally\cite{GeimHallProbe,BruyndoncxAlSquares,ChibotaruAlSquare,MorelleAlShapes,GrigorievaNbDisk,CrenPbIslands2,HaradaAlDisk2,StaleyAlSquares,Morgan-WallWebberBlockade}, with an impressive accumulation of detailed understanding. Studies of vortex motion have also been carried out\cite{KokuboVortexModeLocking,SochnikovNNano,SochnikovPRB,BerdiyorovNbLadder,CaiTwoPeriods} and have led to interesting findings such as vortex-crossing-induced magnetoresistance oscillations \cite{SochnikovNNano,BerdiyorovNbLadder, CaiTwoPeriods} with an amplitude much larger than that of the traditional Little-Parks effect\cite{LittleParks}. Nevertheless, the manipulation of the vortex motion in these mesoscopic devices has yet to be pursued systematically.

In the present work, we use electrical transport measurements to explore the vortex crossing of a mesoscopic type II loop as well as vortex trapping in such a loop. As the magnetic flux threading the loop is varied, the rate of vortex crossing, which is controlled by the free energy barrier through a Boltzmann factor, varies accordingly. The free energy barrier is a function of superfluid velocity, which is a periodic function of the global winding number of the superconducting phase as well as the applied flux\cite{LittleParks}. This leads to a periodically varying rate for vortex crossing\cite{SochnikovNNano,BerdiyorovNbLadder}, and, consequently, a periodically varying magnetoresistance. Interestingly a vortex can also be trapped in the loop under suitable conditions. The trapping of such a vortex, which demands the local phase winding around the vortex core be superimposed on the global phase winding, will have observable effects. It should be emphasized that this Abrikosov vortex trapping is distinct from the so-called ``few-vortex states'' in mesoscopic systems of type I superconductors. In the latter, the ``vortices'' are static solutions to the London or Ginzburg-Landau equations in an applied magnetic field, which are different from Abrikosov vortices in a type II superconductor. For example, as shown in the present study,  the vortex crossing can be directed to constrictions with a minimal crossing barrier and at the same time, Abrikosov vortices can also be trapped in the wide regions of the loop, isolated from the crossing events, providing a clean experimental system for the study of vortex manipulation. 

\begin{figure}
\includegraphics[scale=1]{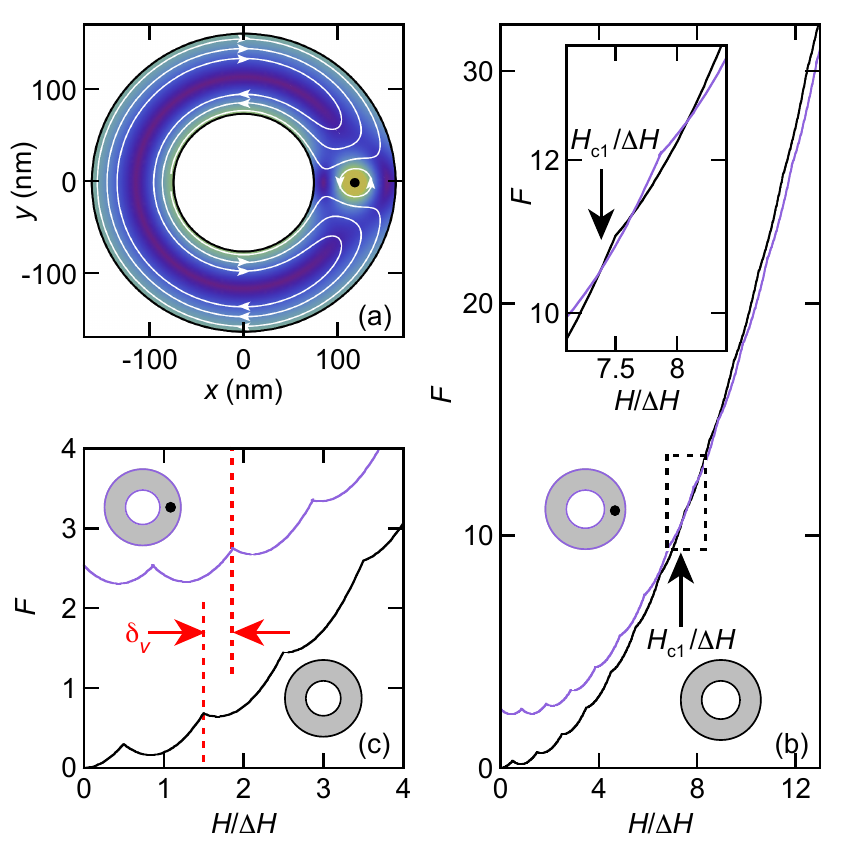}
\caption{(color online) (a) Color plot of current density in a 10~nm thick loop with inner radius $a=73$~nm and outer radius $b=158$~nm (purple (dark gray) is lowest magnitude, yellow (light gray) is highest). Applied field is 0.39~T and a vortex is fixed at ${(x,y)=(115, 0)}$~nm (black disk). Plot generated assuming ${\xi(0)=10}$~nm and ${\lambda(0)=200}$~nm. Loop boundaries are shown in black, and current streamlines are shown in white. (b) Ground state free energy ($\mathcal{F}$) in units of $\epsilon_0$ (Appendix \ref{app-theory}) versus normalized applied magnetic field for loop in (a). Black curves are for vortex-free state, and purple (light gray) curves are with a single vortex trapped at a radius ${v=115}$~nm. Vortex trapping field is indicated. Inset: Close up of boxed region in main panel. (c) Same as (b) over a restricted field range to emphasize the phase shift ($\delta_v$) between vortex-trapped and vortex-free states.}
\label{Theory}
\end{figure}

To obtain the conditions under which Abrikosov vortices can be trapped in a mesosopic superconducting loop of a type II superconductor, we follow the analysis of Kogan, Clem, and Mints\cite{KoganLondon} (see Appendix~\ref{app-theory} for more details) using the London equations in the limit ${d\ll\lambda}$, where $d$ is the loop thickness and $\lambda$ is the magnetic penetration depth. Using the analytic results of these authors, we obtain the current distribution in a superconducting loop with a single vortex trapped at radius $v$, as shown in Figure~\ref{Theory}a. Near the vortex, the current density diverges. Therefore, a cutoff is introduced at $|\vec{r}-\vec{v}|\leq1.2\xi$, where $\xi$ is the superconducting coherence length, to facilitate the plotting. 

Both the free energy ($\mathcal{F}$) of the superconducting loop as a function of global winding number ($N$), vortex position ($v$), and magnetic field ($H$), and the free energy difference between the vortex-trapped and vortex-free states ($V_\text{in}(N,v,H)$) were calculated (Appendix~\ref{app-theory}). In Figure~\ref{Theory}b, we plot the free energy of the trapped vortex state, which depends upon the vortex position. As the magnetic field increases, the free energy of the loop is represented as consecutive parabolas with different $N$. For the vortex-free loop, the transition from state $N-1$ to state $N$ requires the system overcome a free energy barrier with a local maximum at ${H/\Delta H=N-1/2}$. However, when a vortex is trapped in the loop, a different set of parabolas are generated. The backgrounds of the two sets of parabolas, which reflect the kinetic energy resulting from the finite wire width, have different slopes. At the lower critical field, $H_{c1}$, the state with a vortex becomes energetically more favorable than that without one, as shown in Figure~\ref{Theory}b. At this field, $V_\text{in}$ first acquires a global minimum\cite{StanWireHc1}. The background of the free energy originates from the induced currents required by the global phase winding. The crossing of the two curves suggests that introducing a vortex into the loop disrupts this current distribution, leading to the lowering of the free energy of the loop. 

The vortex-trapped parabolas are phase shifted from the vortex-free parabolas by an amount $\delta_v$, as shown in Figure~\ref{Theory}c. To calculate the magnitude of the phase shift, we look for solutions to ${\mathcal{F}(N-1,v,H_N)=\mathcal{F}(N,v,H_N)}$ (Eq.~\ref{Kogan_Fp}) and see that
\begin{equation}
H_N/\Delta H=\left(N-\frac{1}{2}\right)+\frac{\ln(b/v)}{\ln(b/a)}.
\label{shift}
\end{equation}
The acquired phase shift is ${\delta_v=\ln(b/v)/\ln(b/a)}$. It is interesting to note that an apparent free energy oscillation phase shift accompanied by a period change may occur in mesoscopic loops as a result of a field-dependent crossover between effective singly- and doubly-connected geometries\cite{BruyndoncxCrossover}. However, in the mechanism considered in the present work, the phase shift is not accompanied by a change in oscillation period, and therefore corresponds to a different scenario. Vortex trapping discussed here is also distinct from pinning. In vortex pinning, a vortex preferentially occupies a spot where superconductivity is locally suppressed by a material defect. In the scenario presented here, the trapping potential originates from the image force even in the absence of local defects.

While the theoretical prediction is clear, it is not obvious \emph{a priori} that the vortex trapping leads to an experimentally observable effect. For devices recently reported in the literature\cite{SochnikovNNano,SochnikovPRB,BerdiyorovNbLadder}, the measurements appear to have been performed at fields below $H_{c1}$, which depends on the sample size and geometry (Fig.~\ref{VinVsv}). Furthermore, as seen in Eq.~\ref{shift}, $\delta_v$ depends upon the vortex position within the loop, which may not be fixed. Additionally, near the trapping field, the energy difference between the vortex-trapped and vortex-free states is quite small. Indeed, the free energy curves intersect multiple times (inset to Fig.~\ref{Theory}b), potentially prohibiting a clean transition between the vortex-free and vortex-trapped states. So it is quite interesting to see if the phase shift can be detected in a real system.

\begin{figure}
\includegraphics[scale=1]{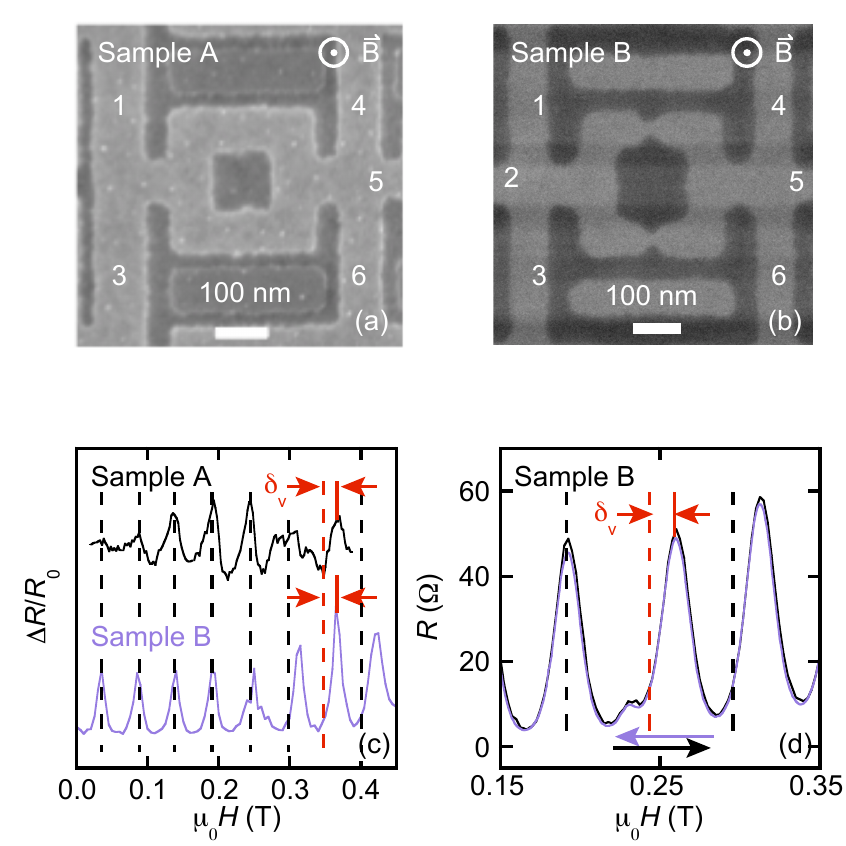}
\caption{(a) Scanning electron microscope (SEM) image of Sample A: a single-crystal NbSe$_2$ square loop with median diameter ${s\approx200}$~nm and arm width $w\approx80$~nm. Current is sourced from 1 to 6 and voltage is measured from 3 to 4. (b) SEM image of Sample B: a NbSe$_2$ loop with $s\approx200$~nm, $w\approx80$~nm, and two artificial constrictions. Current is sourced from 1 to 5 and voltage is measured from 3 to 6. (c) Magnetoresistance of Sample A (black) and Sample B (light purple) at 1.8~K. Sample A data is after subtraction of a smooth background; Sample B data has no subtraction, but is scaled for ease of comparison. Curves offset vertically for clarity. Dashed lines are separated by $\Delta H=520$~Oe, and the phase shift, $\delta_v\approx0.3\Delta H$, is indicated. (d) Magnetoresistance of Sample B at 1.8~K and 900~nA for increasing (black) and decreasing (light purple) magnetic fields. Dashed lines are separated by $\Delta H=520$~Oe.}
\label{characterization}
\end{figure}

Experimentally, we used a combination of electron-beam lithography and CF$_4$ reactive ion plasma etching to prepare mesoscopic loops from atomically thin crystals of NbSe$_2$. The pinning force found in bulk NbSe$_2$ is small. In addition, superconductivity was found to survive down to single-unit-cell thickness\cite{StaleyNbSe2}, making NbSe$_2$ the material of choice for the present study. In the bulk, the superconducting coherence length is $\xi(0)=10$~nm, and the magnetic penetration depth is ${\lambda=200}$~nm\cite{SanchezNbSe2Quantities}. The thickness of our starting crystals is ${\lesssim10}$~nm, so the condition ${d\ll\lambda}$ is satisfied.  The fabrication of NbSe$_2$ devices with feature sizes comparable to the vortex normal core size was described previously\cite{MillsNbSe2Fab}. 

Shown in Figure~\ref{characterization}a is a scanning electron microscope (SEM) image of a NbSe$_2$ square loop (Sample A) with median diameter, $s=200\pm4$~nm, arm width, ${w=80\pm4}$~nm (uncertainty comes from edge roughness and SEM resolution), and thickness, $t=9\pm1.3$~nm (measured by AFM-calibrated color code\cite{StaleyNbSe2,MillsNbSe2Fab}). This square loop, which has the same effective area and arm width as the loop considered in Figure~\ref{Theory}, is easier to fabricate than a circular loop, but likely complicates a direct quantitative comparison to theory. Measurement leads were fabricated on either side of the ring to allow for four-terminal transport measurements. Measurements were carried out using standard current-biased DC techniques under high vacuum ($P\lesssim10^{-5}$~Torr) in a Quantum Design Physical Property Measurement System with a base temperature of 1.8~K and a 9~T superconducting magnet. The magnetic field was oriented perpendicular to the plane of the device as indicated. 

Sample A is fully superconducting with an onset transition temperature of $T_c=6.3$~K, which is only slightly reduced from the bulk value of 7.1~K. The residual resistivity ratio ($RRR\equiv R(300\,\text{K})/R(8\,\text{K})$) is 3.3 for this loop, which is typical for this thickness of NbSe$_2$,\cite{StaleyNbSe2,ElBanaFlakes} indicating the processing did not degrade the quality of the flake. In Figure \ref{characterization}c, we present the magnetoresistance oscillations of Sample A at 1.8~K after removing a smooth background resistance. The oscillations have a period of ${\Delta H\approx520}$~Oe, which is consistent with the measured geometry, and the amplitude of the oscillations is much larger than that expected from the Little-Parks effect, confirming the oscillations originate from vortex crossing. We see that the local maxima in resistance coincide with half-integer values of $H/\Delta H$ (dashed lines) up to $\mu_0H\approx0.25$~T. Above 0.25~T, the oscillations acquire a phase shift, $\delta_v\approx0.3\Delta H$, which is the expected signature of vortex trapping. The field at which this phase shift occurs is in reasonable agreement with the calculated value of $H_{c1}$ (0.38~T from Fig.~\ref{VinVsv}a). Qualitatively similar behavior was observed in multiple samples. We note that the subtraction of a non-monotonic background can in principle introduce an artificial phase shift into otherwise periodic data. However, we can confidently rule out this possibility in the case of Sample A. The observed phase shift between the low-field and high-field peaks is nearly 160~Oe, but we find that the background subtraction shifts the local maxima by less than 25~Oe (see Fig.~\ref{shoulder} and accompanying discussion).

As discussed above, by deliberately fabricating artificial constrictions in the loop, we can direct the vortex crossing to the constrictions with a minimal crossing barrier, and simultaneously isolate the trapped vortex from the crossing events by confining it to the wider regions where $H_{c1}$ is lowest (Fig.~\ref{VinVsv}). An SEM image of a representative NbSe$_2$ loop with a pair of constrictions is shown in Figure~\ref{characterization}b (Sample~B); this device features the same dimensions as Sample~A with the exception of the constrictions. While the addition of the constrictions makes this device in principle a SQUID, we restrict our analysis to the vortex crossing and trapping, rather than quantum interference measurements. In Figure~\ref{characterization}c we plot the magnetoresistance at 1.8~K. In this device, the magnetoresistance oscillations are much more pronounced, with no background subtraction necessary when plotting the data of this sample. This is likely because the vortex crossing events are localized to the constrictions where the energy barrier to crossing is suppressed. Further discussion of the effect of the constrictions on vortex crossing can be found in Appendix~\ref{app-background}, but we wish to focus now on the evidence for vortex trapping. A clear phase shift of $0.3\Delta H$ in the magnetoresistance oscillations is again seen at $+0.25$~T. A number of phase shifts at various positive and negative fields were found (Fig.~\ref{SQUID_RvH}), due either to a sequence of vortex trapping or a change in the location of the trapped vortex. The period of oscillation always remains unchanged, distinguishing this result from the mechanism presented in Ref.~\onlinecite{BruyndoncxCrossover}. We observe no hysteresis in the magnetoresistance (Fig.~\ref{characterization}d), confirming the vortex trapping is not the result of pinning centers, but is instead determined by the field-dependent free energy. Multiple samples were fabricated in the geometry of Sample B, and all show qualitatively similar behavior; namely, pronounced periodic magnetoresistance oscillations which are superimposed on a minimal background resistance and acquire a discrete and reproducible phase shift at a critical magnetic field.

\begin{figure}
\includegraphics[scale=1]{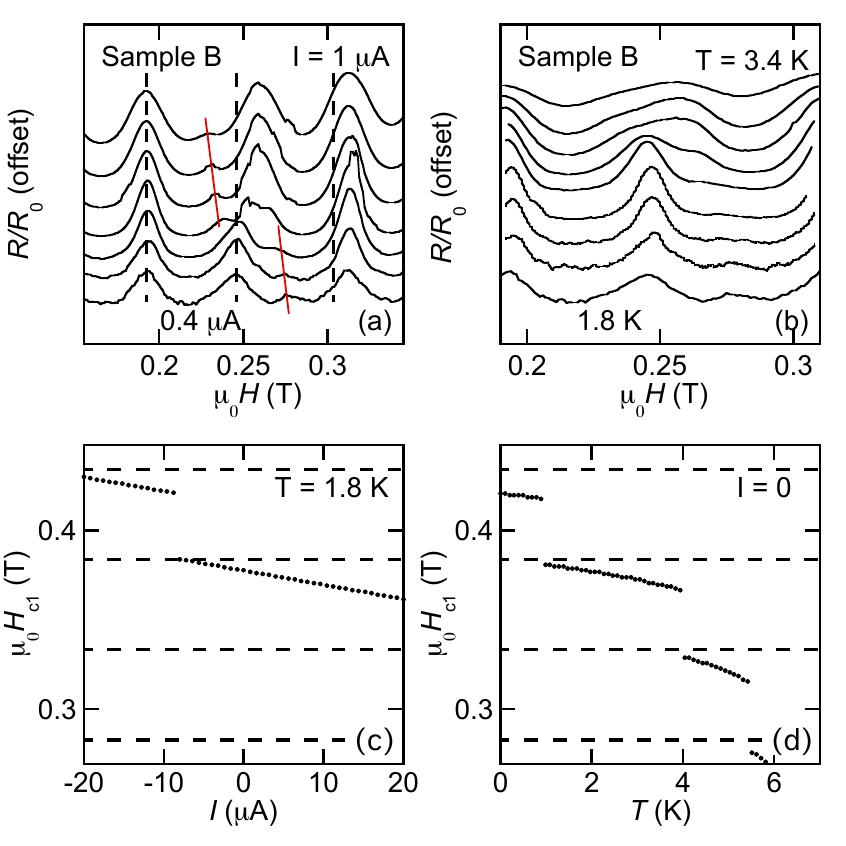}
\caption{(a) Normalized resistance ($R/R_0$) of Sample B versus applied field ($\mu_0H$) at 1.8~K and 0.4, 0.5, 0.6, 0.7, 0.8, 0.9 and 1.0~$\mu$A (bottom to top). Vertical dashed lines indicate the expected (unshifted) positions of peaks 4, 5, and 6. Slanted red lines track approximate value of $H_{c1}$ as indicated by local resistance maxima. Curves offset vertically for clarity. (b) $R/R_0(H)$ for Sample B at 0.4~$\mu$A and 1.8, 2.0, 2.2, 2.4, 2.6, 2.8, 3.0, 3.2, and 3.4~K (bottom to top). Curves offset vertically for clarity. (c) Calculated ${\mu_0H_{c1}}$ versus applied current ($I$) at 1.8~K. Horizontal dashed lines are placed at half-integer multiples of $H/\Delta H$. (d) Calculated ${\mu_0H_{c1}}$ versus temperature ($T$) at $I=0$. Horizontal dashed lines are placed at half-integer multiples of $H/\Delta H$.}
\label{toggle}
\end{figure}

In Figure~\ref{toggle}a we plot the magnetoresistance oscillations of Sample B around 0.25~T at various applied currents. The dashed vertical lines are placed at the expected (unshifted) fields of the $N=4$, 5, and 6 peaks. At low currents, the $N\leq5$ peaks are not phase-shifted, but the $N=6$ peak is (along with subsequent higher peaks not shown), indicating ${4.5\Delta H\leq H_{c1}\leq5.5\Delta H}$. At higher currents, the first phase-shifted peak is the $N=5$ peak, indicating the vortex stability field has decreased to ${3.5\Delta H\leq H_{c1}\leq4.5\Delta H}$. At all currents, the phase shift is equal in magnitude, as demonstrated by the consistent position of the $N=6$ peak. A similar effect is observed when sourcing a fixed current but varying the temperature of the system (Fig.~\ref{toggle}b). The magnetoresistance oscillations broaden with increasing temperature, but it is still clear the first phase-shifted peak changes from $N=6$ to $N=5$ as temperature increases. We also observe an additional local maximum in the magnetoresistance curves which evolves systematically with current (indicated with solid red lines in Fig.~\ref{toggle}a), and may be related to the fine structure near $H_{c1}$ seen in the inset to Figure~\ref{Theory}b. 

The observed features agree with the expectations of the theoretical picture presented above and demonstrate that both temperature and applied current provide additional means of manipulating the trapped vortex. The temperature dependence of $V_\text{in}$ is contained in the $\xi(T)$ term. Assuming that the applied current is uniformly distributed within the loop, an additional Lorentz force on the vortex is expected, resulting in the addition of a linear term to the $V_\text{in}$ curves in Figure~\ref{VinVsv}. Numerical solutions for $H_{c1}$ as a function of applied current and temperature are shown in Figures~\ref{toggle}c-d for the square loop. $H_{c1}(I)$ should be an even function for our system, since in our measurement geometry, current counter propagates in both arms, but in Figure~\ref{toggle}c, we consider $H_{c1}$ in only one arm due to an experimentally observed asymmetry between the arms (see Appendix~\ref{app-asymmetry}).

The discontinuities in the calculated ${H_{c1}}$ curves can be understood by considering the energy difference between the N and N+1 states at a given applied field. ${\mathcal{F}_0(N+1,H)-\mathcal{F}_0(N,H)}$ is periodic in $H$ (Eq.~\ref{Kogan_F0}) and maximal just above half-integer values of $H/\Delta H$ (indicated by dashed lines in Figs.~\ref{toggle}c-d). Going from a vortex-free to vortex-trapped state is similar to increasing the winding number from $N$ to ${N+1}$, and is therefore most energetically costly where ${\mathcal{F}_0(N+1,H)-\mathcal{F}_0(N,H)}$ is maximal. Our measured $R(H,I)$ curves appear to be consistent with a jump in $H_{c1}$. The observed resistance peaks are sufficiently sharp so that both the non-shifted and shifted $N=5$ peaks would be seen at intermediate currents if the critical field for vortex stability evolves smoothly. Given that only one or the other is found experimentally, and that the additional local resistance maximum jumps discontinuously between the 0.6 and 0.7~$\mu$A curves, at least a rapid decrease in $H_{c1}$ from ${H\approx5\Delta H}$ to ${H\lesssim4.5\Delta H}$ is present as $I$ increases, if not an actual discontinuity. 

The manipulation of both vortex crossing and trapping in a mesoscopic loop demonstrated in the present work lays a foundation for further investigations into vortex motion in mesoscopic samples. For example, if the trapped vortices behave as complex quantum mechanical particles, this system may possess macroscopic quantum coherence and form macroscopic energy levels, which can then be measured by radio or microwave measurements\cite{WallraffSingleVortex}. Furthermore, if the trapped vortex can perform coherent quantum motion, detection of the Aharonov-Casher interference of Abrikosov vortices may be attempted\cite{AharonovCasher,ElionAC}. From a technological perspective, the controlled manipulation of Abrikosov vortices within mesoscopic superconducting devices lays the groundwork for a variety of superconducting electronics\cite{VillegasRectifier,HastingsVortexRatchet,BerciuVortexSpintronics}.

Useful discussions with K. Roberts and A. J. Leggett are gratefully acknowledged. The work at Penn State is supported by DOE under Grant number DE-FG02-04ER46159 with nanofabrication done at the Penn State MRI Nanofabrication Lab under NSF Cooperative Agreement 0335765, NNIN with Cornell University as well as DMR 0908700. Work in China was supported by MOST of China (Grant 2012CB927403) and NSFC (Grants 11274229 and 11474198).

\appendix
\section{Current distribution and free energy in the London approximation}
\label{app-theory}

\label{current}
The current density, $\vec{g}$, in a superconducting loop containing a single vortex (+) or anti-vortex (-) at position $\vec{v}$ can be calculated by solving the equations
\begin{equation}
\frac{2\pi\Lambda}{c}\vec{\nabla}\times\vec{g}=\pm\phi_0\delta(\vec{r}-\vec{v})-H,
\label{london}
\end{equation}
\begin{equation}
\vec{\nabla}\cdot\vec{g}=0,
\label{div}
\end{equation}
where $\Lambda(T)=2\lambda(T)^2/d$ is the Pearl length, $\phi_0=h/2e$ is the flux quantum, and $H$ is the applied field. If the radius of the loop, $r$, satisfies $r\ll\Lambda$, as is the case with the devices considered in the main text, the self-fields of the currents within the loop are negligibly small. By introducing a scalar stream function,
\begin{equation}
\vec{g}=\vec{\nabla}\times G\hat{z},
\label{Gdef}
\end{equation}
Eq. \ref{div} is automatically satisfied, and Eq. \ref{london} can be expressed as
\begin{equation}
\frac{2\pi\Lambda}{c}\nabla^2G=\mp\phi_0\delta(\vec{r}-\vec{v})+H.
\label{with-G}
\end{equation}
The linearity of Eq. \ref{with-G} allows for solutions of the form
\begin{equation}
G=G_v+G_H,
\end{equation}
where
\begin{equation}
\nabla^2G_v=\mp\frac{c\phi_0}{2\pi\Lambda}\delta(\vec{r}-\vec{v})
\label{Gv}
\end{equation}
and
\begin{equation}
\nabla^2G_H=\frac{c}{2\pi\Lambda}H.
\label{Gh}
\end{equation}

The latter equation is readily solved, and the former can be solved using techniques from electrostatics. Eq.~\ref{Gv} is analogous to that of an electric charge trapped between two grounded, concentric cylinders, for which the exact solution is given by Jacobi elliptic functions. For loops of the size considered in Ref.~\onlinecite{KoganLondon} as well as in the main text, the exact solution can be approximated by 
\begin{align}
G(r,\theta)&\approx\frac{cH}{8\pi\Lambda}r^2+G_0\ln\frac{r}{a}\nonumber\\
&\pm\frac{c\phi_0}{4\pi^2\Lambda}\text{Re}\left\{\ln\frac{\sin[\pi\ln(vre^{\imath\theta}/a^2)/2\ln(b/a)]}{\sin[\pi\ln(v/re^{\imath\theta})/2\ln(b/a)]}\right\},
\label{complete}
\end{align}
where
\begin{equation}
G_0=-\frac{c\phi_0}{4\pi^2\Lambda}\left[N\pm\frac{\ln{b/v}}{\ln{b/a}}\right],
\end{equation}
and $a$ and $b$ are the inner and outer radii, respectively. Note that Kogan et al. erroneously omit the factor of 2 in the denominator of the logarithm argument in Eq. \ref{complete}. The presence of a vortex leads to an ambiguity in defining the winding number, $N$, so we follow the convention of Ref.~\onlinecite{KoganLondon} that with a vortex present, contours enclosing the annulus hole and the point $\vec{v}$ acquire a phase of $2\pi(N+1)$, whereas contours enclosing just the annulus hole acquire a phase of $2\pi N$. Eqs. \ref{Gdef} and \ref{complete} are used to generate the current distribution shown in Fig. \ref{Theory}a.

Once the exact current distribution is known, the free energy ($\mathcal{F}$) of the loop can be calculated as the sum of the kinetic and magnetic energies. This energy calculation is the main result of Ref.~\onlinecite{KoganLondon}. Kogan et al. obtain
\begin{eqnarray}
\mathcal{F}&(N,v,H)=\epsilon_v(v)+\epsilon_0\left[\left(N+\frac{\ln(b/v)}{\ln(b/a)}\right)^2\right.\notag\\
&-2\left(\frac{H}{\Delta H}\vphantom{\frac{b^2}{a^2}}\right)\left(N+\frac{b^2-v^2}{b^2-a^2}\right)
+\left(\frac{H}{\Delta H}\vphantom{\frac{b^2}{a^2}}\right)^2\chi\left.\vphantom{\left(N\pm\frac{\ln(b/v)}{\ln(b/a)}\right)^2}\right],\label{Kogan_Fp}
\end{eqnarray}
where 
\begin{equation}\epsilon_v(v)=\frac{\phi_0^2}{8\pi^2\Lambda(T)}\ln\left[\frac{2v\ln(b/a)}{\pi\xi(T)}\sin\frac{\pi\ln(v/a)}{\ln(b/a)}\right]\label{vortex_energy}
\end{equation}
is the self-energy of the vortex, ${\epsilon_0=\phi_0^2\ln(b/a)/8\pi^2\Lambda(T)}$ is a characteristic energy scale, $\chi=\frac{b^2/a^2+1}{b^2/a^2-1}\ln(b/a)$ is a geometric factor, and ${\Delta H=2\phi_0 \ln(b/a)/\left(\pi(b^2-a^2)\right)}$. The energy of the vortex-free state is then
\begin{equation}
\mathcal{F}_0(N,H)=\epsilon_0\left(N^2-2N\left(\frac{H}{\Delta H}\right)+\left(\frac{H}{\Delta H}\right)^2\chi\right).\label{Kogan_F0}
\end{equation} 
Eqs. \ref{Kogan_Fp} and \ref{Kogan_F0} are used to generate the curves in Figs.~\ref{Theory}b-c.

\begin{figure}
\includegraphics[scale=1]{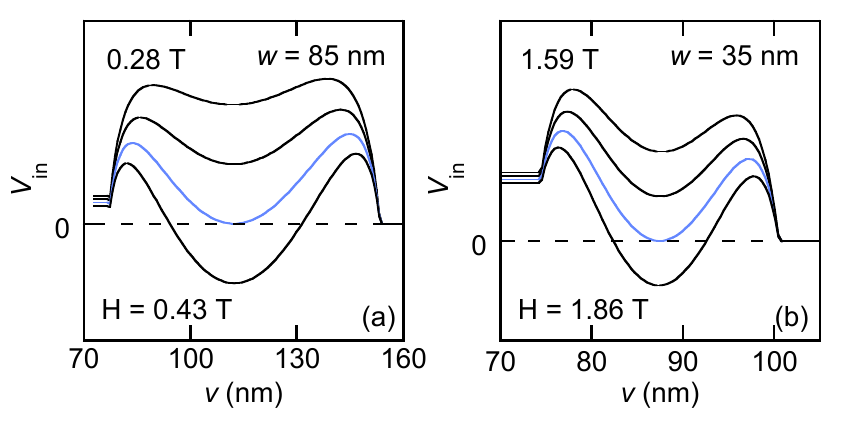}
\caption{(a) Calculated energy ($V_\text{in}$) versus vortex position ($v$) along radius of loop in \ref{Theory}(a) at applied fields of 0.28, 0.33, 0.38, and 0.43~T (top to bottom). Blue curve indicates $H_{c1}$. (b) $V_\text{in}(v)$ for loop with $a=70$~nm and $b=105$~nm at applied fields of 1.59, 1.68, 1.77, and 1.86~T (top to bottom). Blue curve indicates $H_{c1}$.}
\label{VinVsv}
\end{figure}

Finally, the free energy difference between the vortex-trapped and vortex-free states is given by 
\begin{equation}
V_\text{in}(N,v,H)=\mathcal{F}(N,v,H)-\mathcal{F}_0(N,H).
\label{Vin}
\end{equation}
$V_\text{in}$ determines the stability of a vortex at a given position within the arms of the loop. $H_{c1}$ is defined as the field at which a global minimum of $V_\text{in}$ is first found in the sample\cite{StanWireHc1}. In Figure~\ref{VinVsv}a we plot $V_\text{in}$ for the loop in Figure~\ref{Theory}a at different values of applied magnetic field. We see  ${\mu_0H_{c1}\approx0.38}$~T, as indicated by the blue curve. If we consider a thinner loop with $a=70$~nm and $b=105$~nm, we find ${\mu_0H_{c1}\approx1.77}$~T (Fig.~\ref{VinVsv}b).

At this point, we will discuss the limitations of the London approach. The finite size of the vortex core (on the order of $\xi(T)$) is not considered in Eq.~\ref{london}. Thus, in a physical system, the formalism breaks down within $\sim\xi(T)$ of the edges of the sample. To avoid an unphysical divergence, the authors of Ref.~\onlinecite{KoganLondon} set $\epsilon_0=0$ within $\xi(T)/2$ of the sample boundaries. Unfortunately, this essentially arbitrary choice has a direct impact on the numerically calculated $H_{c1}$ when $b-a\sim\xi(T)$. For instance, if we instead choose to set a cutoff of $\xi(T)$, we calculate $\mu_0H_{c1}\approx0.32$~T for the same geometry as considered in Fig.~\ref{Theory}a. For the sake of consistency, we maintain the cutoff employed in Ref.~\onlinecite{KoganLondon}, but the level of quantitative agreement between theory and experiment must be understood in light of these limitations. We stress, however, that the qualitative results reported in the main text, namely, the existence of a free energy oscillation phase shift and the temperature, current, and geometry dependence of the trapping field, are unaffected by the choice of cutoff.

\section{Background resistance and oscillation amplitude}
\label{app-background}

\begin{figure}
\includegraphics[scale=1]{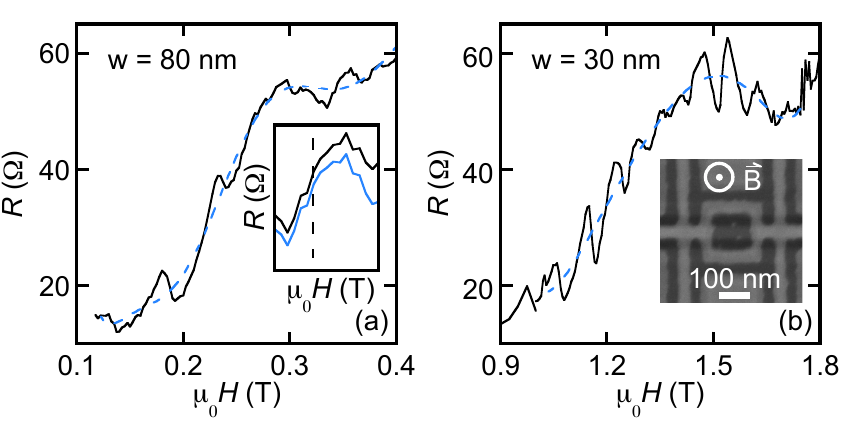}
\caption{(a) Magnetoresistance of thick loop without constrictions. The smooth resistive background indicated by the dashed line is subtracted to obtain the oscillations in Fig.~\ref{characterization}c. The amplitude of oscillations is small compared to the thin loop in (b). Inset: Comparison of raw (black) and background subtracted resistance data (blue) near 0.36~T. Curves shifted vertically for clarity. The dashed line indicates the expected (unshifted) position of the $n=7$ resistance maximum. (b) Magnetoresistance of thin loop without constrictions. A smooth resistive background is still apparent (indicated by the dashed line). The amplitude of oscillations is large compared to the thick loop in (a). Inset: SEM image of thin loop, with $w\approx30$~nm. }
\label{shoulder}
\end{figure}
For loops without artificial constrictions, the observed magnetoresistance oscillations are superimposed on a resistive background. To more clearly observe the vortex-related oscillations and phase shift in Sample A (Figure~\ref{characterization}c), we subtracted a smooth background from the measured resistance data. For completeness, we show the original data for Sample A in Figure~\ref{shoulder}a. The dashed line shows the polynomial fit used for the background subtraction. 

As discussed in the main text, it is possible to introduce an artificial phase shift with an improper background subtraction, especially given that the observed background features a ``shoulder'' very near the critical field for vortex trapping. However, we can be confident that the observed phase shifts in Figure~\ref{characterization}c are not an artifact of a poor background subtraction near this shoulder for three reasons. Firstly, the phase shift is seen in the constricted thick loop without a background subtraction. Secondly, the magnitude of the phase shift is the same for the background subtracted thick loop as for the non-background subtracted constricted thick loop, which would be highly unlikely if the apparent phase shift in the former case was a subtraction artifact. Thirdly, we can directly compare the position of the extrema before and after the background subtraction. In the inset to Figure~\ref{shoulder}a, we plot the raw resistance data (black) and the background-subtracted resistance data (blue) for the thick loop without constrictions. The data has been shifted vertically for ease of comparison. We see that the background subtraction produces no substantial shift in the position of the resistance maximum within the resolution of our measurement. For comparison, the expected location of this maximum in the absence of vortex trapping is indicated by the dashed line. 

In Figure~\ref{shoulder}b, we plot the magnetoresistance data for a second NbSe$_2$ loop. This sample features thin arms with a width $w\approx30$~nm. This arm width was designed to mimic the width of the artificial constrictions introduced to Sample B in the main text. We make two observations regarding this ``thin loop.'' First, the resistance background is still present, which is consistent with our attribution of the background to the lack of a preferred vortex crossing site. Second, the oscillation amplitude is larger than that observed in Sample A (note the same scale of the ordinate axes in Figs.~\ref{shoulder}a and \ref{shoulder}b), which is consistent with the claim that vortex crossing occurs more readily in narrow samples.

Despite all this, our measurements provide only indirect confirmation of the proposed effect of artificial constrictions on the vortex crossing events. It would be beneficial to utilize high-resolution magnetic imaging techniques to confirm the positions of trapped and crossing vortices.

\section{Field asymmetry and multiple vortex trappings}
\label{app-asymmetry}

\begin{figure}
\includegraphics[scale=1]{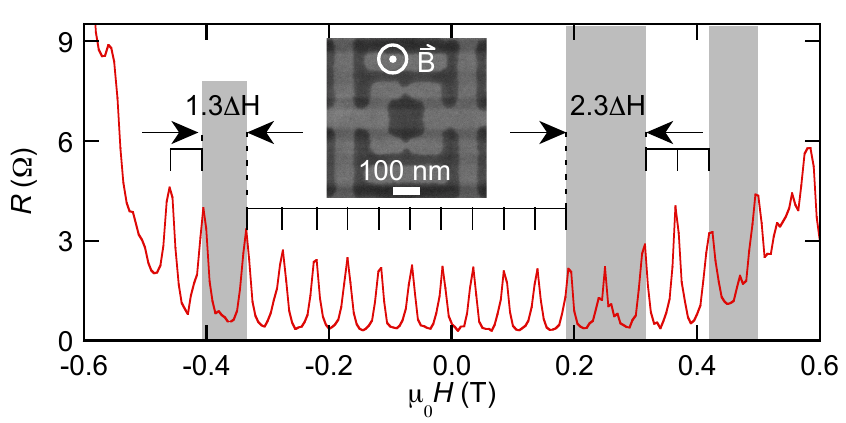}
\caption{Magnetoresistance of thick loop with constrictions at 1.8~K and 650~nA. Vertical lines are separated by $\Delta H=520$~Oe except where indicated. Shaded regions denote where oscillations acquire phase shifts. Inset: SEM image of NbSe$_2$ thick loop with constrictions with $s\approx200$~nm and $w\approx70$~nm. The constrictions on the top and bottom of the loop have a width $\approx30$~nm.}
\label{SQUID_RvH}
\end{figure}

In the case of the loop with constrictions discussed in the main text, several phase shifts are evident at different applied fields (see shaded regions in Fig.~\ref{SQUID_RvH}). These phase shifts may result from subsequent vortex trapping events, though the present theory does not address this situation. The electrostatic analogy employed in Appendix~\ref{app-theory} can be extended to the case of \emph{n} charges equally spaced around two grounded concentric cylinders, though this extension is beyond the scope of this work. Qualitatively, it is reasonable to expect each subsequent vortex trapping event to be accompanied by an additional phase shift in the magnetoresistance oscillations. Alternatively, the additional phase shifts may simply reflect a change in the location of the trapped vortex, since the phase shift depends on $v$ (Eq.~\ref{shift}).

A magnetic field asymmetry is seen in Figure~\ref{SQUID_RvH}, and this likely results from a fabrication-limited device asymmetry. In positive field, the first phase shift occurs at ${\mu_0H\approx0.28}$~T, while in negative field, the first phase shift does not occur until ${\mu_0H\approx-0.38}$~T. However, the phase shift in each orientation is equal in magnitude ($0.3\Delta H$). This asymmetry is not considered in the London calculations, but can be understood as follows: In a perfectly symmetric loop in our measurement geometry, for positive (negative) fields and positive applied current, a vortex will be trapped in the loop at a well-defined $H_{c1}$ in the bottom (top) arm of the loop due to the tilting of the potential energy. The zero-current $H_{c1}$ on either arm is sensitive to inhomogeneities in the sample. This will produce an asymmetry in $H_{c1}$ in the presence of a measurement current. Once a vortex is trapped in an arm, it generates a phase shift of fixed magnitude, as seen experimentally.

\end{document}